%% file: phase_retrieval_notices_arxiv.tex
\documentclass{notices}

\usepackage{amsfonts,amssymb,amsmath,amscd,amsthm}
\usepackage{graphicx}
\usepackage{url}
\usepackage{footnote}
\usepackage{color}
\newtheorem{thm}{Theorem}
\newtheorem{conj}{Conjecture}
\theoremstyle{definition}
\newtheorem{definition}{Definition}
\newtheorem{problem}{Problem}
\newtheorem{example}{Example}
\newcommand{\R}{\mathbb{R}}
\newcommand{\Z}{\mathbb{Z}}
\newcommand{\C}{\mathbb{C}}

\title{
Algebraic theory of phase retrieval
}

\author{
  Tamir Bendory
  \affil{
    Tamir Bendory is a Senior Lecturer at the School of Electrical Engineering of Tel Aviv University. His email address is bendory@tauex.tau.ac.il.
    }
  \and
  Dan Edidin
  \affil{Dan Edidin is Professor of Mathematics at the University of Missouri. His email address is edidind@missouri.edu.}
  }

\begin{document}


\maketitle

\section*{}

The phase retrieval problem first arose in the early 20th century from work on X-ray crystallography. In the last century,  
X-ray crystallography has developed into the leading method to elucidating the atomic structure of molecules, leading to enormous scientific breakthroughs: 
at least 25 Nobel Prizes have been awarded for work directly or indirectly involving crystallography. 
The phase retrieval problem also occurs in numerous other scientific and engineering applications, such as diffraction imaging, ptychography, ultra-short pulse characterization, speech processing, radar, and astronomy. 

In its most general form, this problem can be written as:
\begin{equation} \label{eq:pr}
    \text{find } x\in\Omega \quad \text{subject to} \quad  y = |Ax|^2,
\end{equation}
where $y\in\R^M_{\geq 0}$ is the measurement vector,  $A\in\C^{M\times N}$  is a ``sensing matrix,'' $\Omega$ defines the space of signals of interest, and the absolute value is taken entry-wise. 
Typically, the phaseless measurements are invariant under symmetry groups which depend on $A$ and $\Omega$. Thus, only the orbit of~$x$ under this intrinsic symmetry group can be recovered.

Without model error or noise, namely, when the measurement model~\eqref{eq:pr} is accurate, phase retrieval is a problem of solving real quadratic equations. As such, it is naturally amenable to algebraic techniques, 
including commutative algebra, algebraic geometry, and invariant theory. 
In particular, algebraic methods are powerful techniques for proving that phase retrieval is theoretically possible; i.e., that a particular phase retrieval problem has a unique solution up to the action of its intrinsic symmetry group. 
 Although these methods are typically not algorithmic, they can be used to provide theoretical validations of existing algorithms, and to unveil the fundamental limitations of different methods. 
The purpose of this article is to  discuss recent advances in this growing field of research,  and to publicize open problems that we believe will be of interest to mathematicians in general, and algebraists in particular.

In the rest of the article, we succinctly introduce five specific phase retrieval setups  that we find important and intriguing, present known results, and delineate open mathematical questions. 
We conclude by discussing the limitations of the algebraic point of view, and 
how other mathematical fields, such as information theory, statistics, combinatorics, and optimization, can provide indispensable insights into the phase retrieval problem. 

\subsection*{Phase retrieval with general linear measurements}
The revival of interest in the mathematics of the phase retrieval problem in the last 15 years has emerged from the study of phase retrieval models with a general sensing matrix. 
Specifically, in this setup $A\in\C^{M\times N}$ or $A\in\R^{M\times N}$  is a ``generic'' matrix (usually assumed  to be random), or a
frame\footnote{For the purpose of this article, a frame is a collection of $M$ vectors which span $\R^N$ or $\C^N$.}, with $M>N$, and $\Omega$ is either all of $\R^N$ or $\C^N$. 
While  measurements in practice are not
random, and thus this line of work is of theoretical rather than applicable interest, it attracted the attention of the mathematical community and led to fascinating results in mathematics, statistics, and optimization; see for example~\cite{candes2015phase}. 

For a general real matrix $A \in \R^{M \times N}$,
we can only expect to recover $x \in \R^N$ up to a global sign from
the phaseless measurements $|Ax|$.
Namely,  the intrinsic symmetric group is $\pm 1$.
Likewise, for a general complex matrix, we can recover $x \in \C^N$ only up to multiplication by a scalar $e^{i \theta}\in S^1$, where $S^1$ is the circle group.

For real matrices, there is a remarkably elegant characterization
of when every vector $x \in \R^N$ can be recovered (up to a global sign) from
the phaseless measurements $|Ax|$. To state the result, we introduce the following definition.
\begin{definition} A matrix $A \in \R^{M \times N}$ with row vectors
  $A_1, \ldots , A_M \in \C^N$ satisfies the {\em complement property}
  if, for every subset $S \subset [1,M]$, the vectors $\{A_i\}_{i \in S}$ or
  the vectors $\{A_j\}_{j \in S^C}$
  span $\R^N$.
\end{definition}
  \begin{thm}\cite{balan2006signal}
  	A signal $x\in\R^N$  can be recovered, up to a sign, from $y=|Ax|$ if and only if  $A$ satisfies the complement property.
  \end{thm}
  Note that a necessary condition for a $M \times N$ matrix to satisfy the complement property is $M \geq 2N-1$. This immediately implies that  if $M < 2N-1$, then 
  for any matrix $A$ there  exist a pair of vectors $x,y \in \R^N$
  with $y \neq \pm x$ such that $|Ax| = |Ay|$.

\begin{example}
    The $5 \times 3$ matrix 
    \begin{equation*}
 A = \begin{bmatrix} 1 & 2 &3 \\1 & -1 & 1\\2 & 1 & 4\\1 & 2 & 1 \\2 & -1 & 1 \end{bmatrix},    \end{equation*}
    has full rank but does not satisfy the complement property because the third row  is the sum of the first two rows. 
    In particular,
    if $S = \{1,2,3\}$ then neither $A_1,A_2,A_3$ nor $A_4, A_5$ span $\R^3$.
    Therefore, we know that not all vectors $x$ can be recovered up to a sign from the phaseless measurements $|Ax|$. Indeed, if $x = (1,1,9)^T$ and $y = (19,7,-21)^T$,
then $Ax = (30,9,39,12,10)^T$ and $Ay = (-30,-9,-39,12,10)^T$ so
    $|Ax| = |Ay|$, although $y \neq \pm x$.
  \end{example}
  
  For complex matrices, the situation is more nuanced.
  \begin{thm} \cite{conca2015algebraic} \label{thm.cxphaseretrieval}
   For a generic  complex matrix $A\in\C^{M \times N}$ with $M \geq 4N-4$,   every vector $x \in \C^N$ can be recovered, up to  multiplication
    by a scalar $e^{i \theta} \in S^1$, from the phaseless measurements~$|Ax|$.
  \end{thm}
  Here, we view the space
  of $M \times N$ complex matrices as a real algebraic variety of dimension
  $2MN$. The generic assertion in Theorem \ref{thm.cxphaseretrieval} means
  that the set of matrices for which the conclusion of the theorem holds contains a non-empty open set in the Zariski topology on $(\R^{MN})^2$. However, unlike the case for real matrices, the locus of complex matrices $A$ for which $\C^N/S^1 \stackrel{|Ax|} \to \R^M_{\geq 0}$ is injective is not itself open in the Zariski topology on $(\R^{MN})^2$ because its complement is a semi-algebraic rather than algebraic subset.

  \begin{problem}
  Is there any characterization of an $M \times N$ complex matrix which guarantees that the map
  $\C^N/S^1 \stackrel{|Ax|} \to \R^M_{\geq 0}$
  is injective?
  \end{problem}
  Due to the lack of characterization,  there are relatively few explicit examples of complex $M \times N$ matrices $A$ that guarantee unique recovery of all vectors up to global phase. When $M = 4N-4$, a family of examples was constructed by Bodmann and Hammen in~\cite{bodmann2015stable}. 
   If $N = 2^k +1$ for an integer $k$, a necessary condition for unique  recovery is $M \geq 4N-4$. However, for other values of~$N$, the optimal bound on $M$ is unknown.
 To the authors' knowledge, if $M < 4N-4$, the only examples of $M \times N$ complex matrices which 
 guarantee unique recovery of every signal $x$ (up to an $S^1$ symmetry) are when $M =11$ and $N=4$;  the first examples were constructed by Vinzant \cite{vinzant2015small}.
 
 \begin{problem}
   Determine the optimal bound on $M(N)$ such that, if $M \leq M(N) $, then no signal can be uniquely recovered (up to an $S^1$ symmetry) from $|Ax|.$
   \end{problem}

 One step towards solving this problem was made by Heinosaari, Mazarella and Wolf \cite{heinosaari2013quantum}, 
 who used  results from topology on minimal embeddings of projective spaces
 to prove that  $M(N) \geq 4N-4 -2s_2(N-1)$, where $s_2(N-1)$ is the number of ones in the binary expansion of $N-1$. When $N=4$, this implies $M(N) \geq 8$,
leaving open
the possibility  that Vinzant's construction does not yield the lowest possible bound.

\subsection*{X-ray crystallography}
\begin{figure}
    \centering
    \includegraphics[width=1\linewidth]{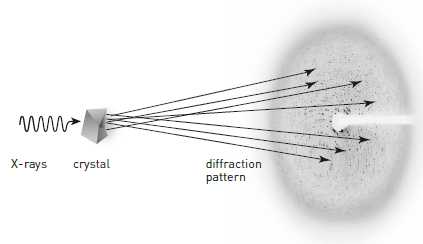}
    \caption{An illustration of an X-ray crystallography experiment.
    The crystalline structure causes an X-ray beam  to diffract into many specific directions. The crystallographic phase retrieval problem is to recover a signal (e.g.,  a molecular structure) from its diffraction pattern. 
\label{fig:Xrayl}}
\end{figure}

X-ray crystallography---a prevalent
technology for determining the three-dimensional atomic structure of molecules---is by far the largest phase retrieval
application.
In X-ray crystallography, the
signal is the electron density function of the crystal --- a periodic arrangement of a repeating,
compactly supported unit
\begin{equation}
    x_c(t) = \sum_{s\in S} x(t-s), 
\end{equation}
where $x$ is the repeated motif and $S$ is a large, but finite, subset of a lattice $\Lambda \subset \R^D$; the dimension $D$ is usually two or three. 
The crystal is illuminated with a beam of X-rays producing a diffraction pattern, which is equal to
\begin{equation}
|\hat x_c(k)|^2=|\hat{s}(k)|^2|\hat{x}(k)|^2,
\end{equation}
where  $\hat{x}$ and $\hat{s}$ are, respectively, the Fourier transforms of the signal $x$ and a Dirac ensemble defined on $S$. 
Figure~\ref{fig:Xrayl} presents an illustration of an X-ray crystallography experiment\footnote{\url{https://www.nobelprize.org/prizes/chemistry/2009/press-release/}.}. 

As the size of the set $S$ grows (the size of the crystal), the support of
the function $\hat{s}$ is more concentrated in the dual lattice $\Lambda^*$. 
Thus, the diffraction pattern is approximately equal to a discrete set of samples of $|\hat{x}|$ on $\Lambda^*$. 
 This
implies that the acquired data are the Fourier magnitudes of a $\Lambda$-periodic signal on $\R^D$ (or,
equivalently, a signal on $\R^D/\Lambda$), defined by its Fourier series
\begin{equation}
    x(t) = \frac{1}{\text{vol}(\Lambda)} \sum_{k\in\Lambda^*} \hat{x}(k)e^{i \langle t,k \rangle}.
\end{equation}
This signal is supported only at the sparsely spread positions of atoms. 
Thus, the crystallographic phase retrieval entails finding a K-sparse signal $x\in\R^N$ satisfying $y=|Fx|^2,$ where $F\in\C^{N\times N}$ is the discrete Fourier transform (DFT) matrix and $K\ll N$.  
 The problem can be equivalently formulated as recovering a K-sparse signal from its periodic
autocorrelation function:
\begin{equation}
a_x[\ell] = \sum_{n=0}^{N-1}x[n]\overline{x[(n+\ell)\bmod N]}.    
\end{equation}
While this article does not focus on algorithmic questions, 
we mention that a set of benchmark problems for evaluating crystallographic phase retrieval algorithms was designed in~\cite{elser2018benchmark}.

In most models, the signal is assumed to be real, so the auto-correlation function is a real quadratic function.
In this case, $a_x[\ell] = a_x[N-\ell]$, so we typically only consider the $\lceil N/2 \rceil$
entries $a_x[0], \ldots , a_x[\lfloor N/2 \rfloor]$
of $a_x$.
Besides  a global sign change, the periodic auto-correlation  is also invariant under cyclic shifts $x[i] \mapsto x[(i+\ell)\bmod N], \, \ell\in\mathbb{Z},$ and  reflection $x[i] \mapsto x[N-i]$. 
{Hence, signal recovery is possible only up to the action  of the group $\{\pm 1\} \times D_{2N}$, where $D_{2N}$ is the dihedral group.}
In particular, for any element $\sigma \in D_{2N}$ and any signal $x$ with support $S \subset [0,N-1]$,   {the signals $\pm \sigma x$ have support $\sigma S$ and the same periodic auto-correlation as~$x$.}

We say that two subsets $S, S' \subset [0,N-1]$ are equivalent if they lie in the same orbit of the dihedral group $D_{2N}$. For a binary signal  (all entries zeros or ones), the auto-correlation is determined by the cyclic difference
multi-set
  $$S-S=\{(b-a) \bmod \pm 1 \mid
  a \leq b \in S\} \subset [0, \lfloor N/2 \rfloor],$$
  where each difference is counted with multiplicity. For example, if $S = \{0,1,2,4\} \subset [0,7]$, then
$S-S = \{0^4, 1^2, 2^2,,3^1, 4^1\}$ and the auto-correlation
vector is $(4,2,2,1,1) \in \R^5$.

The phase retrieval problem for binary sets is equivalent to the combinatorial question of whether two subsets 
with the same cyclic difference multi-sets are dihedrally equivalent. This question does not have an affirmative answer. 
For example, the subsets of
$[0,7]$, $\{0,1,3, 4\}$ and $\{0,1,2,5\}$ both have cyclic difference
multi-sets $\{0^4, 1^2, 2^1, 3^2, 4^1\}$ but are not equivalent.
However, simple numerical experiments show that this phenomenon is quite rare. 
\begin{problem}  For a fixed ratio $|S|/N$, prove that the proportion of  non-equivalent sets with the same difference set is asymptotic to 0 as $N \to \infty$.
\end{problem}

At the other extreme, we can consider a model where  the non-zero entries of the sparse vector are assumed to be arbitrary. In this case, the goal is to prove that a suitably {\em generic} sparse vector is determined up to the action of $\{\pm 1\} \times D_{2N}$ from its periodic auto-correlation function.

In~\cite{bendory2020toward}, we conjectured that if $|S-S| > |S|$, then a generic vector $x$ with support in $S$ is uniquely determined by its periodic auto-correlation up to the action of the group $\{\pm 1\}\times D_{2N}$.
(Here, we view $|S-S|$ as a set rather than a multi-set.)
Verifying this conjecture for a given value of $N$ can be done computationally, although not efficiently, in two steps.

The first step is support recovery; that is, verifying that the auto-correlation determines the support of a generic signal up to {dihedral} equivalence. Note that for a generic signal $x$ with support $|S|$, the support of $a_x$ is the cyclic difference set $|S-S|$. With this observation in hand,  support recovery  can be verified from the following conjecture.

\begin{conj} \label{conj.supportrecover}
If $S,S'$ are two non-equivalent subsets of $[0,N-1]$ of size $K$
with $|S-S|=|S'-S'| \geq  K$, then the incidence variety 
$I_{S,S'} = \{(x,x') \mid a_x = a_{x'}\} \subset L_S \times L_{S'}$
has dimension strictly less than $K$.  (Here $L_S$ and
$L_{S'}$ refer to the subspaces of vectors with support in $S$ and $S'$, respectively.)
\end{conj}

The reason that Conjecture~\ref{conj.supportrecover} implies support recovery
follows from the fact the image of $I_{S,S'}$ under the projection onto the first
factor
$\pi_S \colon I_{S,S'} \stackrel{\pi_S} \to L_S$
is the set of $x \in L_S$ such
that there exists a vector $x' \in L_{S'}$ with $a_x = a_{x'}$. If $\dim I_{S,S'} < K = \dim L_S$, then we know that there is a non-empty Zariski open
set of vectors $x \in L_S$ for which there is no
vector $x' \in L_{S'}$ with the same auto-correlation function. Since
the number of possible subsets $S'$ is finite, affirming Conjecture~\ref{conj.supportrecover}
implies that for a generic vector $x \in L_S$ any vector with the same auto-correlation must have an equivalent support set.

The second step is signal recovery for a signal with known support. 
Let $D_S$ be the subgroup of $D_{2N}$ that preserves a subset $S \subset [0,N-1]$. 
Signal recovery follows from the following conjecture.
\begin{conj} \label{conj.signalrecovery}
If $|S-S| > |S|$ then the incidence variety
$I_S = \{(x,x') | a_x = a_{x'}\} \subset L_S \times L_{S'}$ has dimension $|S|$ and degree $2|D_S|$.
\end{conj}
To see that Conjecture \ref{conj.signalrecovery} implies signal recovery, note that
$I_S$ always contains $2|D_S|$ linear subspaces of dimension $|S|$ consisting of pairs
$\{(x, \pm \sigma x) | \sigma \in D_S\}$. Thus, if $\dim I_S = |S|$ then these linear subspaces must necessarily be irreducible components of $I_S$ of maximal dimension $S$. If in addition $\deg I_S = 2|D_S|$, then these are the only irreducible
components of dimension $S$. It follows that for a generic vector $x \in L_S$, the only vectors with the same auto-correlation as $x$ are the $2|D_S|$ vectors of the form $\pm \sigma x$ for $\sigma \in D_{S}$. This is illustrated further in Example~\ref{example.signalrecovery}.

Although we do not know yet how to prove   Conjecture~\ref{conj.supportrecover} and Conjecture~\ref{conj.signalrecovery}, they can be verified for small values of $N$ using a computer algebra system~\cite{bendory2020toward}.

\begin{example} \label{example.support_recovery.appendix}
The following example verifies a single case
of Conjecture~\ref{conj.supportrecover}. Let $S = \{0,1,2,4\}$ and $S' = \{0,1,2,5\}$ be subsets
	of $[0,7]$. 
	Let 
	\begin{equation*}
	    \begin{split}
	        x &= (x_0, x_1, x_2, 0, x_4, 0, 0, 0) \in L_S, \\
	        y &= (y_0, y_1, y_2, 0, 0, y_5, 0, 0) \in L_{S'}.  
	    \end{split}
	\end{equation*}
By computing the auto-correlations $a_x$ and $a_y$ explicitly, it can be shown that $a_x = a_y$ if and only if the following five equations are
	satisfied:
	\begin{equation} \label{eq.incidence_eqs}
          \begin{array}{lcc}
	x_0^2 + x_1^2  + x_2^2 + x_4^2- y_0^2- y_1^2 - y_2^2 - y_5^2 & = & 0,\\
	x_0x_1 + x_1x_2 -  y_0y_1- y_1y_2 & = & 0,\\
	x_0x_2 + x_2 x_4 - y_0y_2 & = &0,\\
	x_1 x_4 - y_2 y_5 - y_5 y_0 & = & 0,\\
	x_0x_4 - y_1 y_5 & = & 0.
	  \end{array}
          \end{equation}
	Thus, the incidence 
	\begin{equation*}
	 I_{S,S'} = \{(x,y)\,|\, a_x = a_y \} \subset L_S \times L_{S'},   
	\end{equation*}
    is the algebraic subset of $\R^4 \times \R^4$ defined by the set of equations~\eqref{eq.incidence_eqs}. Therefore, the generators of the ideal of	$I_{S,S'}$ are the five polynomials in the left-hand side of~\eqref{eq.incidence_eqs} included 
        in $\R[x_0,x_1,x_2,x_4, y_0,y_1,y_2, y_5]$.
	The Hilbert polynomial of this ideal is $32P_2-80P_1 + 80P_0$, which means that $I_{S,S'}$
	is a 3-dimensional affine algebraic subset of $\R^4 \times \R^4$
	and therefore its image under $\pi_S$ to $\R^4$ has dimension
	at most 3. It follows that for a generic vector $x$
	in the 4-dimensional vector space
	$L_S$, there is no corresponding vector $x' \in L_{S'}$ with the same auto-correlation as $x$.
\end{example}

\begin{example} \label{example.signalrecovery}
We give an example which verifies a case of
Conjecture \ref{conj.signalrecovery}. 
        Let $S = \{0,1,2,5\} \subset [0,7]$ and let
        $L_S$ be subspace of $\R^8$
         with support in $S$. {The set $S$
        	is preserved by the element $\sigma \in D_{16}$ of order two,
        	which is a reflection composed with a shift by two. Thus, $|D_S| =2$ for this subset.}
         If $x=(x_0, x_1, x_2, 0, 0, x_5, 0, 0)$
        and $y = (y_0, y_1, y_2, 0, 0, y_5, 0, 0)$, then
        $a_x = a_y$ if and only if the following  equations are satisfied:
        \begin{equation} \label{eq.incidence_eqs2}
          \begin{array}{lcc}
	x_0^2 + x_1^2  + x_2^2 + x_5^2- y_0^2- y_1^2 - y_2^2 - y_5^2 & = & 0,\\
	x_0x_1 + x_1x_2 -  y_0y_1- y_1y_2 & = & 0,\\
	x_0x_2  - y_0y_2 & = &0,\\
	x_2x_5 + x_5 x_0 - y_2 y_5 - y_5 y_0 & = & 0,\\
	x_1x_5 - y_1 y_5 & = & 0.
	  \end{array}
        \end{equation}
        Let $I$ be the ideal in $\R[x_0,x_1, x_2, x_5, y_0, y_1, y_2, y_5]$
        generated by the five polynomials in the left-hand side of~\eqref{eq.incidence_eqs2}.
        The equations~\eqref{eq.incidence_eqs2} are clearly satisfied 
        if $x = y$ or $x = -y$. Thus, the 4-dimensional linear subspaces
        $L_1 = \{(x,x)\,|\,x \in L_S\}$ and $L_{-1} = \{(x,-x)\,|\, x \in L_S\}$
        are in $Z(I)$, where $Z(I)$ denotes the algebraic subset
        of $L_S \times L_S$ defined by the ideal $I$.
         In addition,  $y = \pm (x_2,x_1,x_0,0,0,x_5,0,0)$
         	are also solutions to  equations~\eqref{eq.incidence_eqs2},
        so there are two additional 4-dimensional linear subspaces
        	$L_\sigma = \{ (x, \sigma x)\,|\, x\in L_S\}$ and $L_{-\sigma} = \{ (x,-\sigma x)\,|\,x \in L_S\}$ in $Z(I)$.

        We calculated the Hilbert polynomial of
        the ideal~$I$ to be $P_I = 4P_3 + 10P_2 -30P_1 + 20P_0$.
        Since $Z(I)$
        contains four linear subspaces
        $L_1, L_{-1}, L_{\sigma}, L_{-\sigma}$, then  
        $ Z(I) \setminus (L_0 \cup L_{-1} \cup L_{\sigma} \cup L_{-\sigma})$, namely,
        \begin{equation} \label{eq:example_known_support}
	\{(x,x') \,|\, a_x = a_{x'} \text{ and } x' \neq g\cdot x \text{ for some } g \in \{\pm 1\} \times D_S\},
\end{equation}
has dimension at most 3. Hence, for generic $x \in L_S$,
if $a_x = a_{x'}$ then $x' = g \cdot x$ for some $g \in \{\pm 1\} \times D_S$. 
\end{example}

Interestingly, using tools from harmonic analysis and information theory, it was recently  proven that  $K$-sparse, symmetric signals are determined uniquely from their periodic auto-correlation
for $K = O(L/\log^5(N))$  for large enough $N$~\cite{ghosh2021multi}.

\paragraph{Future directions.}
Thus far we have discussed 
a binary and generic model for X-ray crystallography.
In practice, however, the model should account for sparse signals whose non-zero entries are taken from a finite (small) alphabet; this alphabet models the relevant type of atoms, such as hydrogen, oxygen, carbon, nitrogen, and so on. Any analysis of this model would be probabilistic, but we expect that in the case of fixed finite alphabet, the probability that a sparse signal can be recovered from its auto-correlation is asymptotic to one as the signal length $N \to \infty$.

\subsection*{Fourier phase retrieval} 
In this section, we consider the problem of recovering
a signal  from its aperiodic auto-correlation:
\begin{equation}
\tilde a_x[\ell] = \sum_{n=0}^{N-1-\ell}x[n]\overline{x[n+\ell}],  
\end{equation}
where $x[n] =0$ if $n \notin [0,N-1]$ and $x[0]$ and $x[N-1]$ are non-zero.
This problem arises in an important imaging technique called coherent
diffractive imaging (CDI); see~\cite{shechtman2015phase} and references therein. 
The aperiodic auto-correlation of a complex signal 
is invariant under the action of the group 
$O(2) = S^1 \ltimes  \{\pm 1\}$, where $S^1$ acts by multiplication by a global constant $e^{i \theta}$ and $-1\in \{\pm 1\}$ acts by conjugation and reflection; these symmetries are typically referred to  as trivial ambiguities. 
Thus, we aim to recover the orbit of $x$ from $\tilde a_x$. 
However,
for generic $x$, there are $2^{N-2}$ orbits with the same aperiodic auto-correlation~\cite{beinert2015ambiguities},  referred to as non-trivial ambiguities. 

To understand the non-trivial ambiguities of recovering a signal from its aperiodic auto-correlation, we can rephrase the problem from the point of view of the Fourier transform.
If we view a signal $x \in \C^N$ as a function
$[0,N-1] \to \C$, then its Fourier transform
\begin{equation}
\hat{x}(\omega) = \sum_{n=0}^{N-1} x[n]\omega^{n},
\end{equation}
 is a polynomial of degree $N-1$ on the unit circle $\omega\in S^1$.
 In the literature, the problem of recovering a signal from the
 Fourier intensity function
\begin{equation}
A_x(\omega) = |\hat{x}(\omega)|^2,
\end{equation}
 is called the Fourier phase retrieval problem. 
Note that $A_x(\omega)$
is a real valued trigonometric polynomial of degree
$2N-1$, which can be expanded as 
$A_x(\omega) = \sum_{\ell = -N}^{N} \tilde a_x[\ell] \omega^\ell$. 
Thus, the function $A_x(\omega)$ encodes equivalent information as the aperiodic auto-correlation.

The absolute value of the discrete Fourier transform 
and the corresponding periodic auto-correlation can be recovered by evaluating $A_x(\omega)$ at the $N$-th roots of unity. 
In particular, $A_x(\omega)$ is equivalent to the information of $|Fz|$, where $z\in\C^{2N}$ so that $z[n]=x[n]$ for $n=0,\ldots,N-1$, and zero otherwise (this is referred to as the support constraint in the phase retrieval literature). 
In this sense, the Fourier phase retrieval is also a special case of~\eqref{eq:pr}.

If we extend the function $\hat{x}(\omega)$ to a polynomial of degree $N-1$ on the entire complex plane, then it has
$N-1$ roots, $\gamma_1, \ldots, \gamma_{N-1}$, and 
we can then write 
\begin{equation}
\hat{x}(\omega) = x_{N-1} \prod_{i=1}^{N-1} (\omega - \gamma_i).
\end{equation}
The following result gives a complete description of the set of vectors
with the same aperiodic auto-correlation as a given vector $x$  and therefore characterizes both the trivial and non-trivial ambiguities in Fourier phase retrieval.
 
\begin{thm}\cite{beinert2015ambiguities} \label{thm.flips}
A vector $x'$ has the same Fourier intensity function $A(\omega)$ 
as $x$ if and only if there is a subset $S \subset [1,N-1]$ and angle $\theta$ such that
\begin{equation}
\hat{x'}(\omega) = e^{i \theta} \prod_{i \in S}
\gamma_i (\omega - \overline{\gamma}_i^{-1})\prod_{j \notin S} (\omega - \gamma_j).
\end{equation}
\end{thm}

In this formulation, if $S = \emptyset$ then 
$x' = e^{i \theta}x$ for some  $e^{i \theta}\in S^1$, and if $S = [1,N-1]$ then $x'$ is a scalar multiple of
the vector obtained from $x$ by reflection and conjugation. It follows that if the roots 
$\gamma_1, \ldots , \gamma_{N-1}$ are distinct and 
$\{\gamma_1, \ldots , \gamma_{N-1}\} \cap \{\overline{\gamma_1}^{-1}
, \ldots , \overline{\gamma_{N-1}}^{-1}\} = \emptyset$  then, modulo
the group $S^1 \ltimes \{\pm 1\}$, there are $2^{N-2}$
vectors with same Fourier intensity function as $x$. For more detail on the ambiguities in one-dimensional Fourier phase retrieval, see~\cite{beinert2015ambiguities, edidin2019geometry}.
\begin{example}
The following vectors
all have the same Fourier intensity function
      \begin{equation*} 
      	A(\omega) =  9/2 \cos(3\theta) + 45/4 \cos(2\theta) + 91/2 \cos \theta + 205/2,
      \end{equation*}
where $\omega = e^{-i \theta}$ is the coordinate
on $S^1$, 
but are unrelated by a trivial
ambiguity:
  $$\begin{array}{ccl}
    x_1 & = & (9/2,9, 1/2,1),\\
    x_2 & = & (3/2,3 +4i, 3/2+ 8i, 3),\\
    x_3 & = & (3/2, 3 - 4i,3/2 - 8i, 3),\\
    x_4 & = &  (9, 9/2, 1, 1/2).
  \end{array}
  $$
       In this example, the roots of $\hat{x}_1(\omega)$ are $\{3i, -3i, -1/2\}$, the roots of $\hat{x}_2(\omega)$ are $\{i/3, -3i, -1/2\}$, the roots of $\hat{x}_3(\omega)$
       are $\{3i, -i/3, -1/2\}$, and the roots of $\hat{x}_4(\omega)$ are $\{3i,-3i, -2\}$.
  \end{example}

Although the Fourier phase retrieval problem is not well-posed, a small amount of additional information is sufficient to recover a signal from
its Fourier intensity function $A_x(\omega)$. For
example, a generic signal~$x$ can be recovered, up to
rotation $e^{i\theta}\in S^1$, from
$A(\omega)$ and {$|x[\ell]|$}
for any $\ell \neq (N-1)/2$ and up to rotation and conjugate reflection from
$|x[(N-1)/2]|$ \cite{beinert2018enforcing}. This and related results play an important role in proving phase retrieval results for short-time Fourier transform (STFT) measurements discussed in the next section. One information-theoretic question about the Fourier intensity function which arises in this context is the following.
\begin{problem}
  Suppose that a subset $S$ of the entries of a generic vector $x \in \C^N$
  are known. What is the fewest number $r$ of values of $A(\omega_1), \ldots , A(\omega_r)$ needed to determine $x$.
\end{problem}
Note that if $S \neq \emptyset$ then $r \leq 2N-1$ since
$A_x(\omega)$ is determined by its values at $2N-1$ distinct angles
and at least one entry of $x$ is known. In \cite{bendory2020blind}, a bound was given based on the size of the difference
set $|S-S|$. However, it is unknown if this bound is optimal.

\paragraph{Higher dimensions.}
To consider the Fourier phase retrieval problem for multi-dimensional signals (say, images), we view a signal in $(\C^N)^d$ as a function 
	$[0,N-1]^d \to \C$, and its $d$-dimensional 
	Fourier transform is the polynomial on the torus $(S^1)^d$:
\begin{equation}
\sum_{(n_1, \ldots , n_d) \in [0,N-1]^d} x[n_1, \ldots , n_d] \omega_1^{n_1} \ldots \omega_d^{n_d}.
\end{equation}
It is well known that if $d > 1$ then almost all signals in $(\C^{N})^d$ can be recovered from the absolute value of the $d$-dimensional Fourier transform or, equivalently, their corresponding aperiodic autocorrelations; this is a direct corollary of the fact that almost all polynomials of degree greater than one, in dimension greater than one, are irreducible over the complex numbers~\cite{hayes1982reducible}.
Nevertheless, it was recently shown, using tools from differential geometry and linear algebra, that 
the problem is, in general, ill-conditioned without additional information about the signal~\cite{barnett2020geometry}.

\subsection*{Periodic short-time Fourier transform and ptychography} 
Ptychography is a computational method of microscopic imaging, in which the specimen is scanned by a localized beam and Fourier magnitudes of overlapping windows are
recorded. 
Mathematically, these are the magnitudes of short-time Fourier transform (STFT) measurements.
The STFT of a signal $x\in\C^N$ can be interpreted as the Fourier transform of the signal multiplied by a sliding window $w\in \C^W$ with
$W \leq N$. Therefore, the phaseless measurements are given by
\begin{equation} \label{eq.stft}
	y[k,r] = \left|\sum_{n = 0}^{N-1} x[n] w[rL -n]e^{-2\pi i nk/N}\right|,
\end{equation}
for  $0\leq k\leq N-1$ and $0\leq r\leq R-1$, 
where $L$ is the separation between sections, 
$R = N/\gcd(N,L)$ is the number of short time sections,  and $w[n] =0$ for $W \leq n \leq N-1$. In this model, the signal and window are assumed to be $N$-periodic so all indices are taken modulo $N$.

Equivalently, the phaseless STFT measurements are the non-negative real vectors
$|F D_0 x|, \ldots , |F {D_{R-1}x}|$, where
$F$ is the DFT matrix, and $D_0,\ldots,D_{R-1}$ are  diagonal matrices
whose non-zero entries are  cyclic shifts of $w$ by~$rL$.
The STFT phase retrieval problem~\eqref{eq.stft} also appears in speech processing.

If the window is known, then the 
STFT measurements $y[k,r]$ are unchanged when $x \in \C^N$ is replaced by $e^{i\theta} x$, so our goal is to recover a signal up to a global phase {$e^{i\theta}\in S^1$} from the phaseless STFT measurements.

Since the number of the phaseless STFT measurements is $NR$,
an important
information-theoretic question is to determine the fewest number
of STFT measurements needed to ensure generic signal recovery.
In~\cite{bendory2021nearoptimal}, it is proved that a generic $x$ can be recovered from
$\sim 4N$ structured STFT measurements. This information-theoretic bound is close to optimal
since the number of real parameters to be recovered is $2N$.

The phaseless STFT model is closely related to the coded diffraction model, where a deterministic sliding window is replaced by a set of random masks~\cite{gross2017improved}. In this case,   it is proved that $\sim N \log N$ measurements are sufficient to recover
a generic signal;  it is still unknown if a signal can be recovered from $\sim N$ coded diffraction measurements.

STFT measurements have a close mathematical relationship to Gabor frames. Given a vector $w \in \C^N$, a Gabor frame generated by $w$ is the collection of $N^2$ vectors 
of the form 
$w_{\ell,p}[n] = w[n+p] \omega^{\ell n}$, 
where $\omega$ is a primitive $N$-th root of unity. This resembles the STFT model
with $L =1$, but the generator~$w$ is typically assumed to be a vector
in $\C^N$ whereas in STFT~$W$ is typically less than~$N$.
In~\cite{bojarovska2016phase}, the authors prove that for a generic, known
generator $w$, every signal can be recovered, up to global phase, from the $N^2$ phaseless
Gabor measurements. 
On the other hand, 
given that a 
generic $(4N -4) \times N$ sensing matrix
allows unique signal recovery (see Theorem~\ref{thm.cxphaseretrieval}), a natural question for future investigation is the following.
\begin{problem} Is a smaller subset of the Gabor frame sufficient for signal recovery (either for all signals or generic signals)?
\end{problem}

\paragraph{Orbit frame phase retrieval.}
The Gabor frame is an example of what we call an orbit frame.
An orbit frame is a frame whose corresponding sensing matrix has the form
\begin{equation*}
A=\begin{bmatrix} FD_0 \\ FD_1 \\ \ldots \\ FD_{r}\end{bmatrix},
\end{equation*}
where $F$ is the DFT matrix, and $D_1, \ldots , D_r$ are diagonal
matrices whose diagonal vectors are obtained from the action of finite, not necessarily abelian, group
on a generating vector $w$. 
It would be interesting to investigate
	conditions (as a function of the group action and $r$) under which  such frames allow recovering signals (either generic signals or all signals) from phaseless samples.

\subsection*{Beyond quadratic equations: blind phaseless STFT  and FROG}
In ptychography the precise structure of the
window might be unknown a priori and thus standard algorithms in the field optimize over the
signal and the window simultaneously. We refer to this model as the blind  STFT problem. In this case, the blind STFT measurements can be viewed as a bilinear
map $\C^N \times \C^W \to \C^{NR}$, $(x,w) \mapsto Y(x,w)$ where
\begin{equation}
Y(x,w)[k,r] = \sum_{n=0}^{N-1}x[n] w[rL -n] e^{-2\pi i nk/N}.
\end{equation}
Because the window is unknown, 
the phaseless functions $|Y(x,w)[k,r]|^2$ are invariant under the bigger group
of ambiguities $S^1 \times (\C^\times)^\alpha \times \Z_R$, where $\alpha = \gcd(L,N)$, 
acting on the set of pairs $(x,w)$ of signals and windows
as follows: $e^{i\theta} \in S^1$ acts by $e^{i \theta}(x,w) = (e^{i \theta }x, e^{i \theta}w)$; $\lambda = (\lambda[0], \ldots , \lambda[\alpha -1])$
acts on $x$ by $x \to (\lambda[0]x[0], \lambda[\overline{1}]x[1],
\ldots , \lambda[\overline{N-1}]x[N-1])$ and on $w$ by
$(\lambda[0]^{-1} w[0], \lambda[\overline{-1}]^{-1} w[1], \ldots , \lambda[
  \overline{-W + 1}]^{-1} w[W-1])$, where $\overline{j}$ denotes the residue of $j$
modulo $\alpha$. Finally, if $\omega$ is an $R$-th root of unity,
then $\omega(x,w) = (x',w')$, where
$x'[n] = \omega^{\lfloor n/\alpha \rfloor} x[n]$ and $w'[n] = \omega^{\lceil n/\alpha \rceil} w[n]$. 
In~\cite{bendory2021nearoptimal}, Cheng and the authors proved that $\sim 4N + 2W$ structured phaseless blind STFT measurements are sufficient to recover a generic pair
$(x,w)$ up to the action of the group of ambiguities. It is not known if a comparable number of random measurements are sufficient to recover generic signals.

Another example of a phase retrieval that involves  polynomials of degree greater than two is technique called  frequency resolved optical gating (FROG) which is used experimentally to chacterize ultra-short laser pulses. The FROG method measures the Fourier magnitude of the product 
 of the signal with a translated version of
itself, for several translations.
Namely, the FROG measurements are given by
\begin{equation} \label{eq.frog}
	y[k,m] = \left|\sum_{n = 0}^{N-1} x[n] x[n+mL]e^{-2\pi i nk/N}\right|^2,
\end{equation}
for $0 \leq k \leq N-1$ and $0\leq M \leq \lfloor N/L \rfloor$, and any out of range indices are taken to be zero.
In \cite{bendory2020signal},  Eldar and the authors considered the problem of recovering a band-limited signal from the FROG measurements. 
A signal  $x \in \C^N$ is
$B$ band-limited if its Fourier transform $\hat{x}$ is supported within a
$B$-element block of $[0,N-1]$. 
If $B \leq \lfloor N/2 \rfloor$, then FROG measurements are invariant under the action of the group $S^1 \times O(2)$. The main result
of \cite{bendory2020signal} states that $\sim 3B$ structured FROG  measurements are sufficient for recovering generic signals, up to the action of the symmetry
group.

\begin{figure}
    \centering
    \includegraphics[width=1\linewidth]{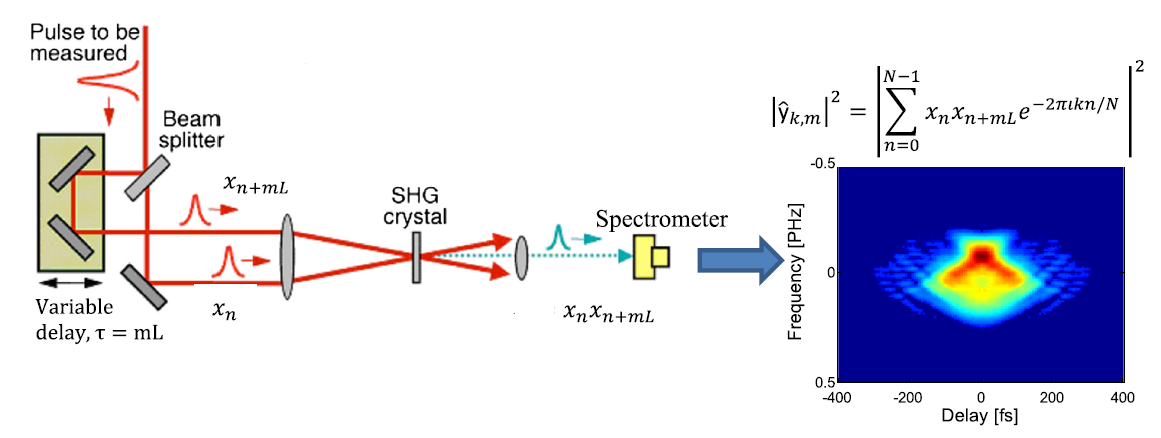}
    \caption{\label{fig:frog} 
    An illustration of the FROG technology for ultra-short pulse characterization~\cite{bendory2020signal}.
}
\end{figure}

\subsection*{Context and limitations}
While algebraic methods are highly effective in understanding the fundamental bounds of different phase retrieval setups, they have their limitations. 
For example, 
in practice the signals in X-ray crystallography 
are not generic, but can be thought of as drawn from a distribution over a finite alphabet;  this alpha-bet represents the set of possible atoms in a biological molecule. Thus, the analysis of this problem requires tools from combinatorics and 
probability theory.

Another prominent 
problem is analyzing the phase retrieval problem in the presence of noise,
where the model~\eqref{eq:pr} does not hold precisely but only approximately; the noise  in many optical imaging applications follows a Poisson distribution.  
In the presence of noise, the goal is estimating the signal to some desired accuracy, rather than solving quadratic equations. 
For such estimation problems, 
algebraic methods can be inadequate, whereas the field of information theory provides the language and the set of tools. 

In addition,  algebraic techniques are typically not very useful in designing and analyzing practical and efficient algorithms.
Based on tools from mathematical optimization and statistics, 
such algorithms were devised for the phase retrieval problem with general linear measurements; see for example~\cite{candes2015phase}. 
Extending these techniques and results to designing provable and efficient algorithms for practical phase retrieval setups, such as X-ray crystallography, ptychography, CDI, and FROG, is a fascinating future research direction 
 at the intersection of mathematics, statistics, optimization, and engineering.

\subsection*{Acknowledgment}
The authors are partially supported by BSF grant no. 2020159.
T.B. is also supported in part by NSF-BSF grant no. 2019752, and ISF grant no. 1924/21. D.E. is also supported by NSF grant no. 19061725.

\input{phase_retrieval_notices_arxiv.bbl}

\end{document}